\documentclass{article}
\usepackage{epsfig}
\newcommand{\bfr}{\begin{flushright}}
\newcommand{\efr}{\end{flushright}}
 
\begin{document}
\title{Charged black holes in a six-dimensional vector-tensor theory 
}
\author{
Kiyoshi Shiraishi\\
Akita Junior College, Shimokitade-sakura 
Akita-shi, Akita 010, Japan
}
\date{in \textit{Festschrift in Honor of Tetsuro Kobayashi's 63rd
Birthday} (Tokyo Metropolitan University, Hachioji, Tokyo, 19 March
1993) edited by H. Minakata, pp.~214--222. (World Scientific,
Singapore, 1994) }
\maketitle
\begin{abstract}
We study a six-dimensional vector-tensor gravitation theory
compactified to $M_4\times S^2$. Here, static spherically-symmetric
solutions to the field equations of the theory are investigated. We
examine the case that the magnetic charge of the ``black hole'' is small
compared with that located in the ``center'' of the extra sphere. The
Hawking temperature of the black hole is calculated in a simple case.
\end{abstract}

It is with great pleasure that I dedicate this paper to Professor
Tetsuro Kobayashi in honor of his 63rd birthday. I wish he'll have more
brilliant years in physics, and hope he will enjoy reading this paper
(though which has nothing to do with experimental physics and
phenomenology!).

\section{Introduction}
It becomes more than one half of a century since Th.~Kaluza presented a
beautiful idea of attributing the gauge symmetry to the symmetry of
compact space \cite{1}. In 80's, many people have quested for the
possibility of identifying the non-Abelian symmetry with the isometry
of the extra spaces \cite{2}.

A difficulty in the program of generating non-Abelian symmetry by a
compact space is the fine-tuning of the four-dimensional cosmological
constant, which arises after compactification. This is due to the
inevitable positive value for the scalar curvature of the extra compact
space with the non-Abelian isometry.

In recent years, many physicists studying unified theories regard that
gauge symmetry as a built-in entity in string theory \cite{3}. The
present authors, however, believe that the non-Abelian Kaluza-Klein
idea is too beautiful to be discarded.

To realize the non-Abelian Kaluza-Klein scenario without the
cosmological constant problem, we suppose, a new action for gravitation
is needed which possesses some kind of higher symmetry than that of
the Einstein-Hilbert action.

The present author \cite{4} and others \cite{5,6,7} have considered the
Gauss-Bonnet action in higher dimensions. It has turned out \cite{6,7}
that there is no cosmological constant problem, moreover, no ordinary
gauge fields and/or gravitons! The Gauss-Bonnet gravity in six
dimensions has an infmite number of massless (extraordinary)
``gravitons'' with mutual interaction if compactified to $M_4\times 
S^2$, where $M_4$ is the four-dimensional Minkowski spacetime and $S^2$
is the extra sphere \cite{7}. The massless fields are expected to be
associated with an infinite-dimensional algebra. The symmetry is
unfortunately too big for our present purpose.
\section{Six-Dimensional Vector-Tensor Theory}
Thus we should examine the model with less symmetry. The present author
and Katsuhiko Yoshida (he also is a student of T.K.) have presented a
new vector-tensor theory in higher dimensions \cite{8}. The action for
the theory is Horndeski's type, i.e., a special, non-minimal
combination of spacetime curvatures and $U(1)$ field strengths
\cite{9}: The Lagrangian density is given by
\begin{eqnarray}
L_G&=&\frac{1}{16}\delta_{EFGH}^{ABCD}F^{EF}F_{AB}R^{GH}_{CD}
\nonumber \\
&=&\frac{1}{4}\left(F^{AB}F_{AB}R-4F^{CB}F_{AB}R^{A}_{C}
+F^{CD}F_{AB}R^{AB}_{CD}\right)\,,
\label{eq1}
\end{eqnarray}
where the totally-antisymmetric Kronecker's symbol is defined by
\begin{equation}
\delta_{EFGH}^{ABCD}=4!\delta_{[E}^{A}\delta_{F}^{B}
\delta_{G}^{C}\delta_{H]}^{D}\,.
\end{equation}

Note that the lagrangian (\ref{eq1}) is linear in the spacetime
curvature. It is not so difficult to see that the $U(1)$ current is
conserved in the theory.

We consider the six-dimensional spacetime $M_4\times S^2$ as a
background spacetime. The kinetic terms for graviton and Yang-Mills
fields appear in four dimensions if we assume magnetic flux on the
extra sphere, a la Randjbar-Daemi et al.\cite{10}. Let us see this.
Consider static Kaluza-Klein ansatz for a background spacetime
\cite{11},
\begin{equation}
d\hat{s}^2=\eta_{\alpha\beta}e^\alpha\otimes e^\beta+b^2\delta_{ab}
(e^a-K^{ia}A^i)\otimes(e^b-K^{jb}A^j)\,,
\end{equation}
where $e^\alpha$ is a vierbein which is expressed by functions of only
four-dimensional  coordinates, while $e^a$ is a zweibein in the
extra sphere which is assumed to be independent of the four-dimensional
coordinates. The constant $b$ is the radius of the extra sphere.
$K^{ia}=K^{im}e_m^a$ are the components of $SU(2)$ Killing vectors
$K^{im}\partial_m$, which satisfy
\begin{equation}
\left[K^i, K^j\right]=\varepsilon^{ij}{}_kK^k\,.
\end{equation}

Similarly, we assume the magnetic flux of the $U(1)$ vector field
\cite{8,10} in addition;
\begin{equation}
\frac{1}{2}F_{AB}e^A\wedge e^B=G e^a\wedge e^b\,,
\end{equation}
where $G$ stands for a magnetic charge.

Substituting these ansatz into the Lagrangian (\ref{eq1}), we get
\begin{equation}
L_G=\frac{G^2}{2b^4}\left(R_{(4)}-\frac{3}{4}b^2K_a^iK^{aj}F^i_{\mu\nu}
F^{\mu\nu j}\right)\,,
\end{equation}
where $R_{(4)}$ is the curvature scalar in the four-dimensional
spacetime and the $SU(2)$ Yang-Mills field strength $F^i_{\mu\nu}$ is
$F^i_{\mu\nu}=\partial_\mu A^i_\nu-\partial_\nu A^i_\mu+
\varepsilon^{ijk}A^j_\mu A^k_\nu$.
Note that the relative sign of the coefficient of Yang-Mills term is
same as that of common Kaluza-Klein theory (derived from the
Einstein-Hilbert action), while the magnitude of the coefficient
differs from usual one. The contribution of the curvature of the extra
sphere is cancelled, so the cosmological constant problem does not
arise.

We have previously considered cosmological evolutions of scale factors
when the scale of the extra space has time-dependence8. The behavior of
the evolution is found to be much alike the one of the five-dimensional
Kaluza-Klein theory and never looks de Sitter-like expansion in the
presence of some normal types of matters \cite{8}.

A major objection to this model of vector-tensor theory is the
ghost-like behavior of the $U(1)$ vector field in four dimensions; a
negative coefficient of a kinematic term for the vector field arises
from the $S^2$ compactification.

In order to mend this point, however, we should not add the usual
Maxwell term in six dimensions; it would induce a four-dimensional
cosmological term if the magnetic flux exists in the extra space
\cite{10}. Compatibility of no cosmological constant and no ghost-like
vector field can be attained by introducing an additional action of
higher order in the field strength for the vector field.

Thus we add the following $O(F^4)$ term in the Lagrangian density:
\begin{eqnarray}
L_V&=&-\frac{\beta^2}{64}\delta_{EFGH}^{ABCD}F^{EF}F_{AB}F^{GH}F_{CD}
\nonumber \\
&=&-\frac{\beta^2}{8}\left\{(F^{AB}F_{AB})^2-2F^A{}_BF^B{}_C
F^C{}_DF^D{}_A
\right\}\,,
\label{eq7}
\end{eqnarray}
Here the constant $\beta$ has a dimension of length.

Again, this Lagrangian ensures the conservation of the current coupled
to the vector field. Further the magnetic monopole configuration still
satisfies the field equation derived from the action. The cosmological
model deduced from the Lagrangian $L_G+L_V$ is unchanged from those
reported in ref.~\cite{8}. Incidentally, the Lagrangian of this type has
been shown to have a huge symmetry in a special limit of the
compactification length \cite{12}. 
\section{Black Holes in Six-dimensional Vector-Tensor Theory}
In this section, we study static solutions in the vector-tensor theory.
As a simplest case, we consider nonsingular, spherically-symmetric,
magnetic black holes. Now, let us find static solutions with spherical
symmetry in the system consisting gravity and the vector field, which
is governed by the Lagrangian $L_G+L_V$.

A static spherically symmetric metric with the extra space $S^2$ can be
written by
\begin{equation}
ds^2=e^{2\varphi(r)}\left[-F^2(r)dt^2+F^{-2}(r)dr^2
+R^2(r)(d\theta^2+\sin^2\theta
d\phi^2)+b_0^2d\tilde{\Omega}^2\right]\,,
\label{eq8}
\end{equation}
where
$d\tilde{\Omega}^2=d\tilde{\theta}^2+\sin^2\tilde{\theta}d\tilde{\phi}^2$
denotes the line element of a sphere with a unit radius. $b_0$ is a
constant, which stands for the asymptotic value of the radius of the
extra sphere. We propose the following asymptotic behavior of $F$, $R$,
and $b$ at spatial infinity:
\[
F(r)\rightarrow 1\,,\quad
R(r)\rightarrow	1\,,\quad
\mbox{and}\quad
\varphi(r)\rightarrow 0\quad\mbox{at}\quad r\rightarrow\infty\,.
\]

We chose the block diagonal form in (\ref{eq8}), and then $SU(2)$
Yang-Mills fields which arise in four dimensions are set to zero in the
present analysis.

For the $U(1)$ magnetic field strengths, we make the ansatz
\begin{equation}
\frac{1}{2}F_{MN}dx^M\wedge dx^N=g \sin\theta d\theta\wedge d\phi
+G\sin\tilde{\theta}d\tilde{\theta}\wedge d\tilde{\phi}\,,
\label{eq9}
\end{equation}
where $g$ and $G$ are magnetic charges of the ``black hole'' and of the
extra sphere, respectively.

The reason for we take the ansatz (\ref{eq8}) with the conformal factor
is to simplify the equation of motion. Similar conformal
transformations of four-dimensional metric are known to make the
leading term of the gravitational Lagrangian in four dimensions the
Einstein-Hilbert lagrangian \cite{13,14,15}.

To get a positive sign in front of the kinetic term for the vector
field in four dimensions, we require $\beta^2>b_0^2$. In this model, the
Newton constant in four dimensions is expressed as
$G_N=b_0^2/(8\pi G^2)$.

The field equations for the metric (\ref{eq8}) and the monopole ansatz
(\ref{eq9}) are
\begin{eqnarray}
& &2\frac{R''}{R}+\left(\frac{R'}{R}\right)^2+2\frac{F'}{F}
\frac{R'}{R}-\frac{1}{F^2R^2}+3(\varphi')^2
\nonumber \\
& &=\frac{(g/G)^2b_0^4}{R^4}\left\{
2\frac{R''}{R}-6\left(\frac{R'}{R}\right)^2+2\frac{F'}{F}
\frac{R'}{R}-\frac{1}{b_0^2F^2}-3(\varphi')^2-6\frac{R'}{R}\varphi'
\right\}\nonumber \\
& &\quad-\frac{\beta^2g^2}{F^2R^4}e^{-2\varphi}\,,
\label{eq10}\\
& &\frac{R''}{R}+\frac{F''}{F}+\left(\frac{F'}{F}\right)^2+2\frac{F'}{F}
\frac{R'}{R}+3(\varphi')^2
\nonumber \\
& &=\frac{(g/G)^2b_0^4}{R^4}\left\{
\frac{F''}{F}+3\varphi''+3(\varphi')^2+\left(\frac{F'}{F}\right)^2
+6\frac{F'}{F}\varphi'-\frac{1}{b_0^2F^2}
\right\}\nonumber \\
& &\quad+\frac{\beta^2g^2}{F^2R^4}e^{-2\varphi}\,,
\label{eq11}\\
& &-3\varphi''-6\left(\frac{F'}{F}+\frac{R'}{R}\right)\varphi'
\nonumber \\
& &=\frac{(g/G)^2b_0^4}{R^4}\left\{
3\frac{R''}{R}-9\left(\frac{R'}{R}\right)^2+6\frac{F'}{F}
\frac{R'}{R}+3\varphi''+6\left(\frac{F'}{F}-\frac{R'}{R}\right)\varphi'
\right\}\nonumber \\
& &\quad+\frac{\beta^2g^2}{F^2R^4}e^{-2\varphi}\,,
\label{eq12}
\end{eqnarray}
where where the prime denotes the derivative with respect to $r$.

We will solve Eqs.~(\ref{eq10}--\ref{eq12}) with a perturbation
expansion in
$(g/G)^2\equiv\gamma^2$, which is assumed to be small. This expansion is
equivalent to the expansion in $b_0^2$, up to the first order we will
treat.

At zeroth order, the solution is given by
\begin{eqnarray}
&
&F_0^2(r)=\left(1-\frac{r_+}{r}\right)\left(1-\frac{r_-}{r}
\right)^{1/2}\,,
\label{eq13a}\\
& &R_0(r)=r\left(1-\frac{r_-}{r}
\right)^{1/4}\,,
\label{eq13b}\\
& &	e^{\varphi_0(r)}=\left(1-\frac{r_-}{r}
\right)^{-1/4}\,,
\label{eq13c}
\end{eqnarray}
where $r_+$ is an integration constant and
$r_-=\frac{4\beta^2g^2}{3r_+}$. The horizon is located at $r=r_+$.

To first order in $\gamma^2$, we choose 
\begin{eqnarray}
& &F=F_0(1+\gamma^2\mu(r))\,,\\
& &R=R_0(1+\gamma^2\rho(r))\,,\\
& &e^\varphi=e^{\varphi_0}(1+\gamma^2\eta(r))\,.
\end{eqnarray}

We find the solution at the first order and large $r$:
\begin{eqnarray}
& &\mu(r)=b_0^2\left(\frac{1}{(2r_-+r_+)r}+\frac{r_-}{(2r_-+r_+)r^2}
+\frac{r_-}{4r^3}\right)+O(r^{-4})\,,\\
& &\rho(r) =b_0^2\left(-\frac{1}{(2r_-+r_+)r}-\frac{r_-}{(2r_-+r_+)r^2}
-\frac{r_-}{4r^3}\right)+O(r^{-4})\,,\\
& &\eta(r)=b_0^2\left(\frac{1}{(2r_-+r_+)r}+\frac{r_+}{2(2r_-+r_+)r^2}
+\frac{r_+}{3(2r_-+r_+)r^3}\right)+O(r^{-4})\,.
\end{eqnarray}

Unlike the analysis of the black holes in string theory with higher
derivatives \cite{14,15,16}, the first order functions $\mu$, $\rho$
and $\eta$ cannot be written in closed forms because the zeroth order
solution has a non-polynomial form.

Now, let us examine the mass of the black hole. At zeroth order, the
metric can be written as
\begin{equation}
ds_0^2=-\left(1-\frac{r_+}{r}\right)dt^2+\frac{dr^2}{\left(1-\frac{r_+}{r}
\right)\left(1-\frac{r_-}{r}\right)}+r^2(d\theta^2+\sin^2\theta d\phi^2)
+b_0^2\left(1-\frac{r_-}{r}\right)^{-1/2}d\tilde{\Omega}^2\,.
\end{equation}

From this metric, one can see the increase of the scale of the compact
space near the black hole. If we adopt this form of the metric, the
zeroth-order gravitational and inertial mass of the black hole
are\begin{eqnarray}
M_{G0}&=&r_+/2\,,\\
M_{I0}&=&(r_++r_-)/2\,,
\end{eqnarray}
respectively, where we set the four-dimensional Newton constant $G_N$
equals to one.
\section{Masses and Temperatures of the Black Holes}
Then in this frame, we find the gravitational mass of the black hole up
to the first order in $\gamma^2$, considering the asymptotic behavior of
$g_{tt}$,
\begin{equation}
M_G=\frac{r_+}{2}\left(1-\frac{4\gamma^2b_0^2}{(2r_-+r_+)r_+}\right)
=M_{G0}\left(1-\frac{\gamma^2b_0^2}{M_{G0}^2+\frac{2}{3}\beta^2g^2}
\right)\,.
\end{equation}

On the other hand, considering the asymptotic behavior of $g_{rr}$ , one
finds the inertial mass of the black hole up to the first order in
$\gamma^2$ is unchanged
\begin{equation}
M_I=M_{I0}\,.
\end{equation}

Instead, if we adopt $\bar{g}_{\mu\nu}=e^{-2\varphi}g_{\mu\nu}$ as the
four-dimensional metric, we see the masses of the black hole by the two
definitions coincide with each other. At zeroth order, we find
\begin{equation}
\bar{M}_{G0}=\bar{M}_{I0}\equiv\bar{M}_0=\frac{1}{2}
\left(r_++\frac{r_-}{2}\,.
\right)
\end{equation}

At first order, we find
\begin{eqnarray}
& &\bar{M}_G=\bar{M}_I\equiv\bar{M}=\frac{1}{2}\left(r_++\frac{r_-}{2}
\right)\left(1-
\frac{4\gamma^2b_0^2}{(2r_-+r_+)(r_-+2r_+)}\right)\nonumber \\
& &=\bar{M}_0\left(1-
\frac{\gamma^2b_0^2}{\left(5\bar{M}_0-3\sqrt{\bar{M}_0^2-
\frac{2}{3}\beta^2g^2}\right)\bar{M}_0}\right)<\bar{M}_0\,.
\end{eqnarray}

Even if $\beta^2g^2$ takes a small value (as long as $\beta^2>b_0^2$ is
satisfied), the masses are modified at the first order in $\gamma^2$. In
such a case, which is equivalent to $r_-\ll r_+$, one finds
\begin{eqnarray}
& &M_{G0}=M_{I0}\equiv M_0\,, \\
& &M_G=M_0\left(1-\frac{\gamma^2b_0^2}{M_0^2}\right)<M_0\,,\\
& &M_I=M_0\,,\\
& &\bar{M}_{G0}=\bar{M}_{I0}\equiv\bar{M}_0\,,\\
& &\bar{M}_{G}=\bar{M}_{I}=\bar{M}=\bar{M}_0
\left(1-\frac{\gamma^2b_0^2}{2\bar{M}_0^2}\right)<\bar{M}_0\,.
\end{eqnarray}

The corresponding Hawking temperature is independent of the conformal
factor, say $e^{2\varphi}$, when it is expressed in terms of $r_+$ and
$r_-$. It can be derived from the regularity at $r=r_+$ in euclidean
spacetime with periodic time. If we further assume $r_-\ll r_+$, we
obtain at first order:
\begin{equation}
T=\frac{1}{4\pi r_+}\left(1+2\gamma^2\mu(r_+)\right)
=\frac{1}{4\pi r_+}\left(1+\frac{2\gamma^2b_0^2}{r_+^2}\right)\,.
\end{equation}

Therefore, the Hawking temperature is expressed in terms of the masses
in each metric as follows:
\begin{equation}
T=\frac{1}{8\pi M_G}\left(1-\frac{\gamma^2b_0^2}{2M_G^2}\right)
=\frac{1}{8\pi M_I}\left(1-\frac{\gamma^2b_0^2}{2M_I^2}\right)
=\frac{1}{8\pi \bar{M}}\,.
\end{equation}

We prefer the choice of	$\bar{g}_{\mu\nu}$ as the physical metric, for
the equivalence principle holds in the primitive way, i.e.,
$\bar{M}_G=\bar{M}_I$. Then the relation between Hawking temperature
and mass is unchanged when the correction of first order in $\gamma^2$
is included, provided that
$r_-\ll r_+$.
\section{Summary}
To summarise: Magnetic black holes in our model of the vector-tensor
theory in six dimension have similar properties as the charged
dilatonic black holes in string theory \cite{14,15,16}, in which
higher-derivative couplings and the dilaton exist.

Since $SU(2)$ gauge fields appears in four dimensions in our model, the
future subject to study is the investigation of the non-Abelian charge
of the black holes in the vector-tensor model. The study on non-linear
interactions of ``electromagnetic waves'' in the compactified model is
to be studied.
\section*{Acknowledgements}
The present author wishes to express his cordial thanks to Professor
Tetsuro Kobayashi for continuous encouragement and support. He also
thanks H.~Minakata and members of the department of physics at TMU for
warm hospitality.

\end{document}